\documentclass[sigconf]{acmart}
\usepackage{multirow} 

\usepackage{subfigure}
\usepackage{booktabs}
\usepackage{tabularx}
\usepackage{makecell}
\usepackage{caption}
\usepackage{xurl}
\usepackage[normalem]{ulem}
\usepackage{amsmath}
\usepackage{soul}
\usepackage{array}
\usepackage{geometry}

\usepackage{tikz}
\usepackage{colortbl}
\usepackage{algorithm}
\usepackage{algorithmic}
\usepackage{enumitem}
\usepackage{arydshln}
\usepackage{stfloats}

\definecolor{lightgreen}{RGB}{173, 216, 230}
\definecolor{lightred}{RGB}{255, 218, 185}

\AtBeginDocument{%
  }

\setcopyright{acmlicensed}
\copyrightyear{2026}
\acmYear{2026}
\acmDOI{xxxxxxx.xxxxxxx}

\acmConference[SIGKDD '26]{Proceedings of the 32nd SIGKDD Conference on Knowledge Discovery and Data Mining}{August 9-13, 2026}{Jeju, Korea}
\acmISBN{979-x-xxxx-xxxx-x/xxxx/xx}

\begin{document}



\title{LISRec: Modeling User Preferences with Learned Item Shortcuts for Sequential Recommendation}



\author{Haidong Xin}
\affiliation{
  \institution{Northeastern University}
  \city{Shenyang}
  \country{China}
}
\email{xinhaidong@stumail.neu.edu.cn}

\author{Zhenghao Liu}
\authornote{\ \ indicates corresponding author.}
\affiliation{
  \institution{Northeastern University}
  \city{Shenyang}
  \country{China}
}
\email{liuzhenghao@mail.neu.edu.cn}

\author{Sen Mei}
\affiliation{
  \institution{Tsinghua University}
  \city{Beijing}
  \country{China}
}
\email{meisen2025@gmail.com}

\author{Yukun Yan}
\affiliation{
  \institution{Tsinghua University}
  \city{Beijing}
  \country{China}
}
\email{yanyk.thu@gmail.com}

\author{Shi Yu}
\affiliation{
  \institution{Tsinghua University}
  \city{Beijing}
  \country{China}
}
\email{yus21@mails.tsinghua.edu.cn}

\author{Shuo Wang}
\affiliation{
  \institution{Tsinghua University}
  \city{Beijing}
  \country{China}
}
\email{wangshuo.thu@gmail.com}

\author{Zulong Chen}
\affiliation{
  \institution{Alibaba Group}
  \city{Hangzhou}
  \country{China}
}
\email{zulong.czl@alibaba-inc.com}

\author{Yu Gu}
\affiliation{
  \institution{Northeastern University}
  \city{Shenyang}
  \country{China}
}
\email{guyu@mail.neu.edu.cn}

\author{Ge Yu}
\affiliation{
  \institution{Northeastern University}
  \city{Shenyang}
  \country{China}
}
\email{yuge@mail.neu.edu.cn}

\author{Chenyan Xiong}
\affiliation{
  \institution{Carnegie Mellon University}
  \city{Pittsburgh}
  \country{United States}
}
\email{cx@cmu.edu}

\renewcommand{\shortauthors}{Haidong Xin et al.}


\def\method{LISRec}
\begin{abstract}
User-item interaction histories are pivotal for sequential recommendation systems but often include noise, such as unintended clicks or actions that fail to reflect genuine user preferences. To address this, we propose Learned Item Shortcuts for Sequential Recommendation (\method{}), a novel framework that explicitly captures stable preferences by extracting personalized semantic shortcuts from historical interactions. \method{} first learns task-agnostic semantic representations to assess item similarities, then constructs a personalized semantic graph over all user-interacted items. By identifying the maximal semantic connectivity subset within this graph, \method{} selects the most representative items as semantic shortcuts to guide user preference modeling. This focused representation filters out irrelevant actions while preserving the diversity of genuine interests. Experimental results on the Yelp and Amazon Product datasets illustrate that \method{} achieves a 13\% improvement over baseline recommendation models, showing its effectiveness in capturing stable user interests. Further analysis indicates that shortcut-based histories better capture user preferences, making more accurate and relevant recommendations. All codes and datasets are available at \url{https://github.com/NEUIR/LISRec}.
\end{abstract}





\maketitle

\begin{figure}[!t]        
\centering
\includegraphics[width=0.47\textwidth]{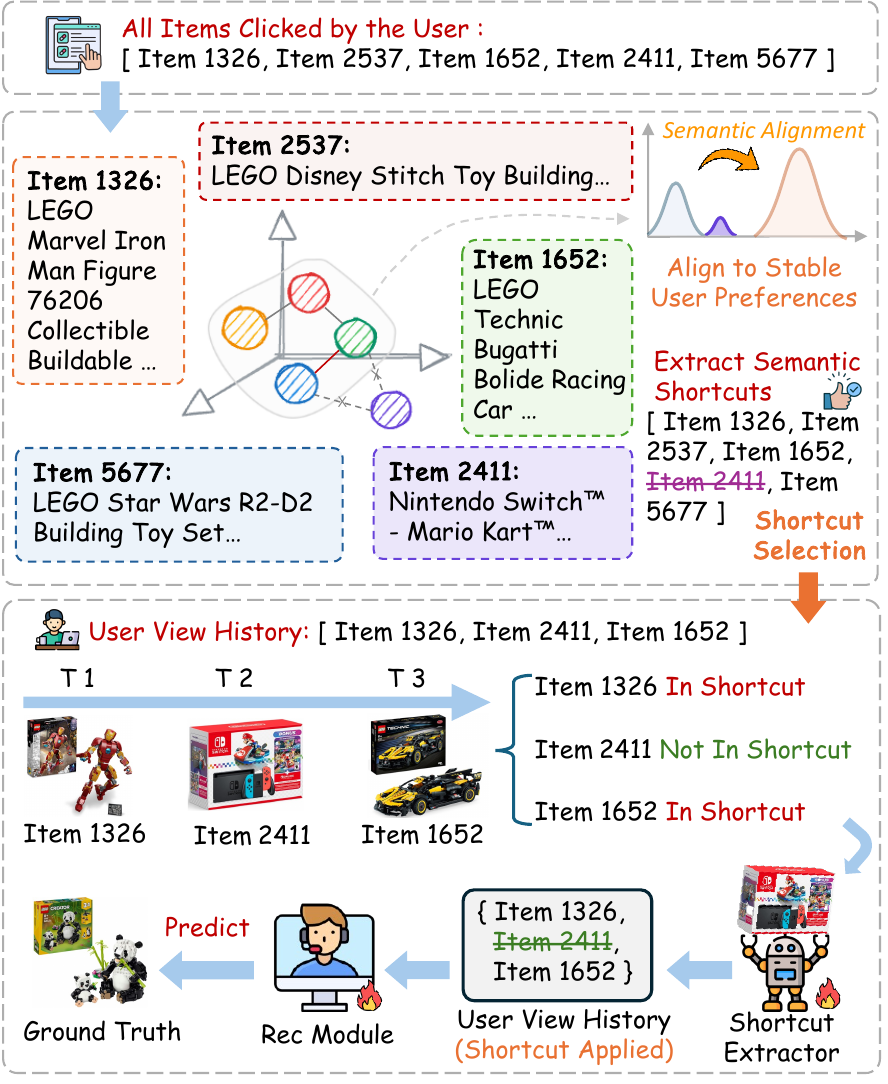}
\caption{\label{fig:intro}Illustration of Our \method{} Method. \method{} constructs a personalized semantic interaction graph and extracts shortcuts, representing dominant item subsets that provide denoised user representations for preference modeling.}
\vspace{-2em}
\end{figure}
\section{Introduction}
Sequential recommendation systems play a pivotal role in modern digital platforms by dynamically aligning recommendations with evolving user preferences, thereby satisfying user needs and mitigating information overload~\cite{wang2019sequential,fang2020deep}. While existing approaches predominantly focus on modeling user behavior through interaction histories, a critical challenge remains unaddressed: real-world interaction sequences inherently contain noise from incidental clicks that diverge from user stable preferences. As shown in Figure~\ref{fig:intro}, a user consistently purchasing LEGO products might occasionally click on unrelated items like gaming consoles due to momentary curiosity. Such noisy interactions dilute the signal of genuine preferences, leading to suboptimal recommendation results.

Current methodologies primarily follow two paradigms: item-ID based modeling~\cite{Wang2018self,Sun2019Bert4rec} and item-content based modeling~\cite{Hou2022uni,liu2023text}. While these approaches learn representations from interaction sequences, they indiscriminately incorporate all items during training, implicitly assuming uniform relevance across interactions. This oversight amplifies noise propagation, as models struggle to distinguish between core preferences and transient outliers~\cite{lin2023self,zhang2024ssdrec}. Recent attempts to address noise either adjust item weights~\cite{wang2021denoising}, or rely on implicit contrastive learning~\cite{yang2023debiased}, but these methods lack explicit mechanisms to identify and eliminate noise, limiting their interpretability.

To address these challenges, we propose Learned Item Shortcuts for Sequential Recommendation (\method{}), a novel framework that explicitly captures stable user preferences by extracting personalized semantic shortcuts from interaction histories. Specifically, \method{} first learns task-agnostic semantic representations to estimate similarities between items, and then constructs a personalized semantic graph using all items previously clicked by the user, based on pairwise item similarities. Next, \method{} identifies a Maximal Semantic Connectivity subset of the personalized semantic graph that best captures the user's underlying preferences, serving as semantic shortcuts to guide the user behavior modeling for recommendation. By focusing on these highly representative items, \method{} effectively filters out irrelevant interactions while preserving the diversity of the user's genuine interests.

Our experimental results on the Amazon Product and Yelp datasets demonstrate that \method{} achieves up to 13\% improvement over a range of baseline models. Our framework generalizes well across different recommendation architectures, highlighting its flexibility. Further analysis shows that the extracted shortcuts consist of semantically similar items, and that these subsets are more predictive of user preferences in next-item recommendation. Overall, \method{} bridges the gap between semantic representation and noise filtering, providing a principled and generalizable solution for improving sequential recommendation performance.
\section{Related Work}
Sequential recommendation systems aim to model user behaviors by leveraging user-item interaction histories. Early studies rely on the Markov Chain assumption~\cite{he2016fusing} and Matrix Factorization techniques~\cite{Rendle2010Factor} for item prediction. To better represent user preferences, recent works have primarily focused on ID-based modeling, which encodes the randomly initialized ID embeddings of user-interacted items using various neural architectures. These architectures include Recurrent Neural Networks (RNNs)~\cite{Hidasi2015session,liu2016context,donkers2017sequential,yang2017neural}, Convolutional Neural Networks (CNNs)~\cite{Jiaxi2018Personalized,Fajie2018asc,yan2019cosrec}, Graph Neural Networks (GNNs)~\cite{Chang2021graph,zhang2022dynamic,cai2023lightgcl}, and self-attention mechanisms~\cite{Wang2018self,Sun2019Bert4rec,du2023frequency,yang2023collaborative}.

Instead of representing items using IDs, existing methods focus more on item content-based recommendation systems, leveraging Pretrained Language Models (PLMs) or Large Language Models (LLMs) to enhance item representations by learning textual semantics from their side information~\cite{Ding2021zero,yuan2023go,liu2023text,chen2024hllm}. These approaches~\cite{liu2023text,chen2024hllm} encode the textual representations of items and further train language models using user-item interaction signals. TASTE~\cite{liu2023text} verbalizes both items and user interaction histories, utilizing the T5 model~\cite{raffel2020exploring} to align user and item representations. Leveraging the emergent abilities of LLMs~\cite{huang2023towards,lin2023can}, HLLM~\cite{chen2024hllm} introduces a hierarchical architecture for sequential recommendation comprising the item LLM and the user LLM.

\begin{figure*}[t]
    \centering
    \includegraphics[width=\textwidth]{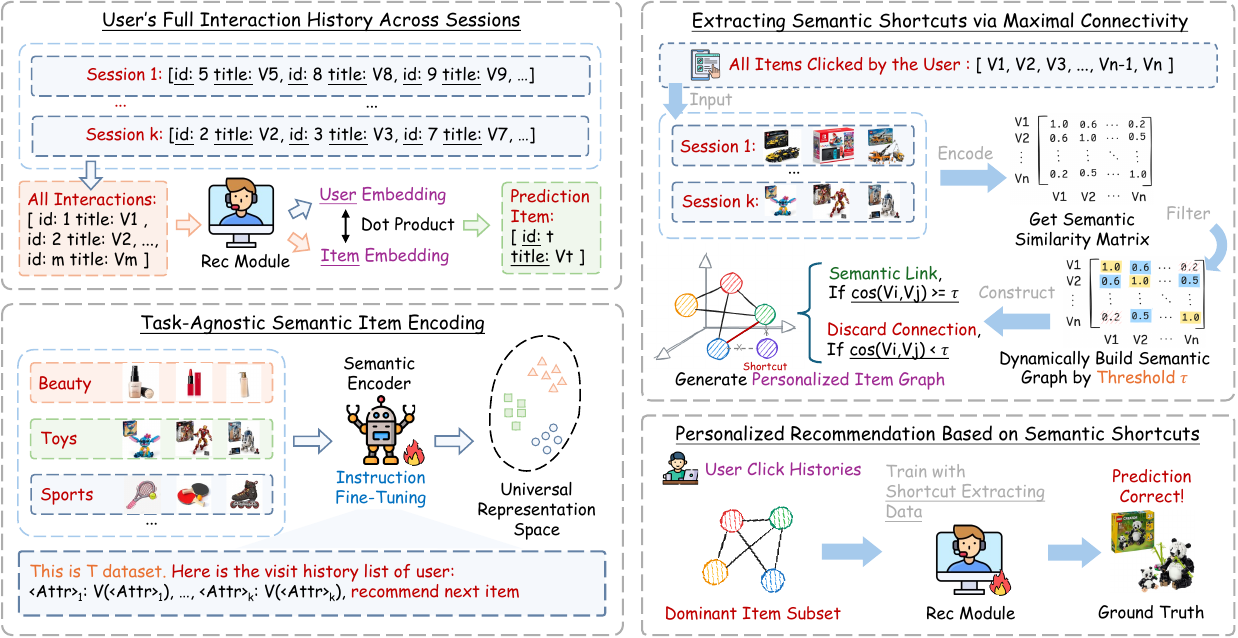} 
    \caption{\label{fig:model}The Model Architecture of \method{}.}
\end{figure*}
User interaction sequences in recommendation systems often contain noise, such as accidental clicks or short-term behaviors that do not reflect users’ true preferences~\cite{lin2023self}. To address this, several methods reduce the influence of noisy interactions by monitoring training dynamics (e.g., high prediction loss or unstable gradients), including adaptive sample reweighting~\cite{wang2021denoising}, label correction via double correction~\cite{he2024double}, and user-specific denoising strategies~\cite{zhang2025personalized}. Beyond loss-based methods, self-supervised denoising approaches~\cite{lin2023self,zhang2024ssdrec} learn item dependencies to identify noisy signals, while other works adjust item weights based on estimated importance~\cite{zhou2022filter,chen2022denoising,yang2023debiased}. However, these approaches typically treat denoising as an implicit process without directly removing noisy interactions.

Another line of research~\cite{zhang2023denoising,zhang2022hierarchical,sun2021does,qin2021world} focuses on explicitly filtering out noise by identifying and removing low-quality items from interaction sequences. These methods predominantly rely on item ID-based representations, which require extensive user-item interaction histories to be effective~\cite{lam2008addressing,schein2002methods,pan2019warm,zhang2021language}. Consequently, they often overlook the potential of text-based semantic information for identifying preference-consistent behaviors~\cite{liu2023text,yuan2023go}.

\section{Methodology}
This section introduces \method{}, a framework for robustly modeling user preferences by extracting semantic shortcuts in a latent space. As shown in Figure~\ref{fig:model}, we first describe the overall architecture of \method{}, which identifies preference-consistent shortcuts to construct denoised user representations (Sec.~\ref{model:preliminary}). We then detail the shortcut extraction process, which identifies the Maximal Semantic Connectivity within each user's personalized item graph (Sec.~\ref{model:graph}).

\subsection{The Framework of \method{}}\label{model:preliminary}
The sequential recommendation task aims to predict the item $v_t$ that satisfies the preference of the user $\mathcal{U}$ at the $t$-th time step according to the user-item interaction history $\mathcal{H}$:
\begin{equation}
\small
\mathcal{H} \rightsquigarrow M_\text{Rec} \rightsquigarrow v_t,
\end{equation}
where the user-item interaction history is represented as $\mathcal{H}=\{v_1,v_2,\dots,v_{t-1} \}$. The recommendation model $M_\text{Rec}$ estimates the relevance between the user-item interaction history $\mathcal{H}$ and an item $v$ by calculating the ranking probability $P(v|\mathcal{H})$: 
\begin{equation}\label{eq:prob}
\small
P(v|\mathcal{H}) = \text{Softmax}_{v} (\vec{\mathcal{H}} \cdot \vec{v}),
\end{equation}
where $\cdot$ denotes the dot product operation. Then we follow~\citet{liu2023text} to get the representations $\vec{v}$ and $\vec{\mathcal{H}}$ for the candidate item $v$ and the user-item interaction history $\mathcal{H}$.

\subsubsection{Item Representation.} 
To fully use the semantic information of items, we verbalize the item $v$ using its identifier and $k$ associated attributes $\texttt{<Attr>}$ through the following template:
\begin{equation}\label{eq:item}
\small
X(v)= \texttt{id:} v(\texttt{id}) \dots \texttt{<Attr>}_k: v(\texttt{<Attr>}_k),
\end{equation}
where $\texttt{<Attr>}_k$ is the name of the $k$-th attribute, while $v(\texttt{id})$ and $v(\texttt{<Attr>}_i)$ correspond to the text descriptions of the item identifier and the $i$-th attribute, respectively. Then we encode each item $v$ into an embedding $\vec{v}$ using the T5 model~\cite{raffel2020exploring}:
\begin{equation}\label{eq:encode}
\small
\vec{v} = \text{T5}(X(v)).
\end{equation}

\subsubsection{Modeling User Preferences via Shortcut Extraction} Users' interaction histories ($\mathcal{H}$) often contain a mixture of long-term interests, short-term preferences, and incidental click behaviors. To model user intent using truly essential items, \method{} tries to model preference-consistent behaviors by building semantic shortcuts in the user-interacted item trajectory to omit some noisy items.

To identify these shortcuts, we introduce a semantic shortcut extractor $M_\text{Filter}$, which selectively retrieves a subset of dominant interactions from the full history $\mathcal{H}$. These shortcuts, denoted as $\mathcal{H}_\text{Shortcut}$, form a condensed yet highly informative sketch of user preferences in the latent semantic space. This distilled representation is then passed to the recommendation module $M_\text{Rec}$ for more effective user behavior modeling. The user representation can be formalized as:
\begin{equation}
\small
\mathcal{H} \xrightarrow{M_\text{Filter}} \mathcal{H}_\text{Shortcut},
\end{equation}
where $\mathcal{H}_\text{Shortcut} \subseteq \mathcal{H}$ denotes the extracted shortcut subset, containing interactions that are semantically aligned with the user's underlying intent. The semantic shortcut extractor $M_\text{Filter}$ are detailed in Section~\ref{model:graph}.

To construct the user representation, we first verbalize the extracted shortcut set $\mathcal{H}_\text{Shortcut}$ by concatenating the textual representations (Eq.~\ref{eq:item}) of the selected items:
\begin{equation}\label{eq:seqtext}
\small
X(\mathcal{H}_\text{Shortcut}) = \mathop{\Vert}_{v_i \in \mathcal{H}_\text{Shortcut}} X(v_i),
\end{equation}
where $\Vert$ denotes the concatenation operation over the subset of items in $\mathcal{H}_\text{Shortcut}$, which may be non-contiguous and unordered with respect to the original history. The concatenated text is then encoded using the T5 model to obtain a semantic representation of the user’s preference:
\begin{equation}\label{eq:seqencode}
\small
\vec{\mathcal{H}}_\text{Shortcut} = \text{T5}(X(\mathcal{H}_\text{Shortcut})).
\end{equation}

\subsection{Extracting Preference Shortcuts via Maximal Semantic Connectivity}\label{model:graph} To robustly model stable user preferences from noisy interaction histories $\mathcal{H}$, we introduce a semantic shortcut extractor $M_\text{Filter}$, which identifies the most representative subset of interactions for preference modeling. Specifically, \method{} first encodes items into a task-agnostic semantic space to estimate pairwise item similarity, then constructs a personalized semantic graph for each user and extracts the maximally connected component as the preference-consistent shortcut. This process effectively filters out noisy or inconsistent behaviors during user representation construction.

\subsubsection{Task-Agnostic Semantic Item Encoding} To obtain semantically meaningful item representations suitable for preference-level reasoning, we train a language model on item textual descriptions using the instruction tuning based item representation method.

For item semantic representation, each item $v$ is represented purely through its descriptive attributes—excluding any item ID—to improve generalization and reduce dependency on sparse or domain-specific identifiers:
\begin{equation}
\small
\Tilde{X}(v)'= \texttt{<Attr>}_1: v(\texttt{<Attr>}_1)\ \dots\ \texttt{<Attr>}_k: v(\texttt{<Attr>}_k).
\end{equation}
Building upon the Next Item Prediction (NIP) objective~\cite{Sun2019Bert4rec}, we serialize each user history $\mathcal{H} = \{v_1, \dots, v_{t-1}\}$ into a concatenated text sequence:
\begin{equation}
\small
\Tilde{X}'(\mathcal{H})= \Tilde{X}(v_{t-1})'; \dots; \Tilde{X}(v_1)'.
\end{equation}
This sequence is passed to the encoder alongside an instruction prompt: ``This is the $\mathcal{T}$ dataset. Here is the visit history list of the user: $\Tilde{X}'(\mathcal{H})$, recommend next item''. The encoder is trained to predict the next item $v_t$, encouraging it to learn semantic dependencies that align with user preferences. After training, the encoder produces item embeddings $\vec{v}$, where pairwise semantic similarity can be evaluated via the cosine similarity:
\begin{equation}\label{eq:sij}
\small
S_{ij} = \cos(\vec{v}_i, \vec{v}_j).
\end{equation}
This representation approach is fully text-driven, enabling the item representation model to operate across item vocabularies and different tasks while capturing semantic regularities grounded in language.

\subsubsection{Modeling Stable Preferences with Personal Item Graphs}\label{model:denoising} To extract the user’s core preference signals from noisy interaction histories, we construct a personalized semantic interaction graph for each user $\mathcal{U}$, where edges capture semantic coherence between interacted items:
\begin{equation}
\small
     G_\mathcal{U} = (\mathcal{V}, \mathcal{E}),
\end{equation}
where $\mathcal{V}=\{v_1,\dots,v_n\}$ includes all items interacted by user $\mathcal{U}$ in all previous sessions, and $\mathcal{E}$ denotes the edge set encoding semantic connections. An undirected edge $e_{ij} \in \mathcal{E}$ is created between items $v_i$ and $v_j$ if their embedding similarity exceeds a threshold $\tau$:
\begin{equation}\label{eq:edge}
\small
e_{ij} =
\begin{cases} 
0 & \text{if } S_{ij} < \tau, \\
1 & \text{if } S_{ij} \geq \tau,
\end{cases}
\end{equation}
where $e_{ij}=0$ is the cosine similarity between the embeddings of $v_i$ and $v_j$ are not connected (Eq.~\ref{eq:sij}). This design naturally filters out semantically dissimilar and outlier items with respect to user preferences.

To identify the user’s dominant preference signal, we extract the largest connected component from the graph $G_\mathcal{U}$:
\begin{equation}
\label{eq:gmax}
\small
C_\text{max} = \arg\max_{C_i} |\mathcal{V}(C_i)|,
\end{equation}
and define the corresponding item subset as the preference shortcut:
\begin{equation}
\label{eq:idenoised}
\small
\mathcal{H}_\text{Shortcut} = \{v_j \mid v_j \in \mathcal{H} \land v_j \in \mathcal{V}(C_\text{max}) \}.
\end{equation}

This shortcut subset serves as a semantically consistent and compact proxy for the user's intent. By selectively modeling only the maximally connected region in the semantic graph, we avoid overfitting to noisy interactions while mitigating over, ensuring richer and more robust preference representations for downstream recommendation.


\section{Experimental Methodology}
This section describes the datasets, evaluation metrics, baseline models, and implementation details used in our experiments.

\textbf{Datasets.} We utilize four sequential recommendation datasets--Beauty, Sports, and Toys from Amazon Product~\cite{mcauley2015image}, and Yelp--to train and evaluate both the sequential recommendation module and the semantic shortcut extractor. More details are shown in Appendix~\ref{app:data}.

\textbf{Evaluation Metrics.} We adopt Recall@10/20 and NDCG@10/20 to evaluate the recommendation performance, which is consistent with prior works~\cite{Xie2022DIF-SR,liu2023text,chen2024hllm}. Statistical significance is determined using a permutation test with $P<0.05$. 

\textbf{Baselines.} We compare \method{} with several widely used methods for sequential recommendation, including vanilla sequential recommendation models and some item denoising techniques.

\textit{Vanilla Sequential Recommendation Models.} We compare \method{} with both ID-based and content-based recommendation models. First, \method{} is evaluated against seven ID-based recommendation models. GRU4Rec~\cite{Hidasi2015session} employs Recurrent Neural Networks (RNNs) to model user-item interaction sequences, while SASRec~\cite{Wang2018self} and Bert4Rec~\cite{Sun2019Bert4rec} leverage self-attention to capture user preferences from sequential interactions. Additionally, we examine models that incorporate item-side information. S$^3$Rec~\cite{Zhou2020s3} enhances self-attention modules through four self-supervised pretraining strategies to better model relationships among attributes, items, and users. ICAI-SR~\cite{yuan2021icai} represents item-attribute relationships using a heterogeneous graph to improve relevance modeling, whereas NOVA~\cite{Liu2021nova} integrates item attributes as side information via attention modules. DIF-SR~\cite{Xie2022DIF-SR} introduces a non-invasive attention mechanism to fuse item attribute information into user behavior modeling effectively. 

Furthermore, \method{} is compared with four content-based recommendation models. CoWPiRec~\cite{yang2023collaborative} optimizes Pretrained Language Models (PLMs) to enhance item representation by aligning it with collaborative filtering representations derived from a word graph. TASTE~\cite{liu2023text} leverages T5~\cite{raffel2020exploring} to directly encode the text representations of both users and items for matching items with user intentions. DWSRec~\cite{zhang2024dual} leverages both fully whitened and relaxed whitened item representations as dual views for effective recommendations. Lastly, HLLM~\cite{chen2024hllm} employs Large Language Models (LLMs) to generate item representations, subsequently utilizing a hierarchical architecture to predict the next item.

\begin{table*}[t]
\centering
\small
\caption{\label{tab:overall}Overall Performance of \method{}. The \textbf{best} results are marked in bold, while the \underline{second-best} results are underlined. The HLLM$^*$ model is reproduced using a 1B-scale LLM. ${\dagger}$ and ${\ddagger}$ denote statistically significant improvements over $\text{DIF-SR}^{\dagger}$ and  $\text{TASTE}^{\ddagger}$, respectively.}
\resizebox{\textwidth}{!}{
\begin{tabular}{l|c|ccccccc|ccccc}
\hline 
\multirow{2}{*}{\textbf{Dataset}} & \multirow{2}{*}{\textbf{Metrics}} & \multicolumn{7}{c|}{\textbf{Item ID based Recommendation Models}} & \multicolumn{5}{c}{\textbf{Text based Recommendation Models}}\\ 
 &  & \textbf{GRU4Rec} & \textbf{Bert4Rec} & \textbf{SASRec} & \textbf{S$^3$Rec}& \textbf{NOVA} & \textbf{ICAI-SR} & \textbf{DIF-SR} & \textbf{CoWPiRec} & \textbf{DWSRec} & \textbf{HLLM$^*$} & \textbf{TASTE} & \textbf{\method{}} \\ 
\hline
\multirow{4}{*}{Beauty} & R@10 & 0.0530 & 0.0529 & 0.0828 & 0.0868 & 0.0887 &0.0879 & 0.0908 & 0.0837 & 0.0989 & 0.0887 & \underline{0.1030} & \textbf{0.1249}$\text{}^{\dagger \ddagger}$ \\
& R@20 & 0.0839 & 0.0815 & 0.1197 & 0.1236 & 0.1237 & 0.1231 & 0.1284 & 0.1280 & 0.1496 & 0.1238 & \underline{0.1550} & \textbf{0.1774}$\text{}^{\dagger \ddagger}$ \\
& N@10 & 0.0266 & 0.0237 & 0.0371 & 0.0439 & 0.0439 & 0.0439 & 0.0446 & 0.0405 & 0.0470 & 0.0499 & \underline{0.0517} & \textbf{0.0641}$\text{}^{\dagger \ddagger}$ \\
& N@20 & 0.0344 & 0.0309 & 0.0464 & 0.0531 & 0.0527 & 0.0528 & 0.0541 & 0.0517 & 0.0597 & 0.0588 & \underline{0.0649} & \textbf{0.0772}$\text{}^{\dagger \ddagger}$ \\
\hline
\multirow{4}{*}{Sports} & R@10 & 0.0312 &0.0295 & 0.0526 & 0.0517 & 0.0534 & 0.0527 & 0.0556 & 0.0588 & 0.0629 & 0.0555 & \underline{0.0633} & \textbf{0.0721}$\text{}^{\dagger \ddagger}$ \\
& R@20 & 0.0482 &0.0465 &0.0773 &0.0758 &0.0759 &0.0762 & 0.0800 & 0.0892 & 0.0958 & 0.0816 & \underline{0.0964} & \textbf{0.1082}$\text{}^{\dagger \ddagger}$ \\
& N@10 & 0.0157 & 0.0130 & 0.0233 & 0.0249 & 0.0250 & 0.0243 & 0.0264 & 0.0300 & 0.0313 & 0.0295 & \underline{0.0338} & \textbf{0.0372}$\text{}^{\dagger \ddagger}$ \\
& N@20 & 0.0200 & 0.0173 & 0.0295 & 0.0310 & 0.0307 & 0.0302 & 0.0325 & 0.0376 & 0.0396 & 0.0360 & \underline{0.0421} & \textbf{0.0463}$\text{}^{\dagger \ddagger}$ \\
\hline
\multirow{4}{*}{Toys} & R@10 & 0.0370 & 0.0533 & 0.0831 & 0.0967 & 0.0978 & 0.0972 & 0.1013 & 0.0502 & 0.0967 & 0.0928 & \underline{0.1232} & \textbf{0.1304}$\text{}^{\dagger \ddagger}$ \\
& R@20 & 0.0588 & 0.0787 & 0.1168 & 0.1349 & 0.1322 & 0.1303 & 0.1382 & 0.0739 & 0.1472 & 0.1306 & \underline{0.1789} & \textbf{0.1896}$\text{}^{\dagger \ddagger}$ \\
& N@10 & 0.0184 & 0.0234 & 0.0375 & 0.0475 & 0.0480 & 0.0478 & 0.0504 & 0.0272 & 0.0457 & 0.0520 & \textbf{0.0640} & \underline{0.0637}$\text{}^{\dagger}$ \\
& N@20 & 0.0239 & 0.0297 & 0.0460 & 0.0571 & 0.0567 & 0.0561 & 0.0597& 0.0331 & 0.0585 & 0.0616 & \underline{0.0780} & \textbf{0.0786}$\text{}^{\dagger}$ \\ \hline
\multirow{4}{*}{Yelp} & R@10 & 0.0361 & 0.0524 & 0.0650  &0.0589 & 0.0681 &0.0663 & 0.0698 & 0.0657 & 0.0702 & 0.0659 & \underline{0.0738} & \textbf{0.0749}$\text{}^{\dagger \ddagger}$ \\
& R@20 & 0.0592 & 0.0756 & 0.0928 & 0.0902 & 0.0964 & 0.0940 & 0.1003 & 0.0959 & 0.1103 & 0.0955 & \underline{0.1156} & \textbf{0.1173}$\text{}^{\dagger \ddagger}$ \\
& N@10  & 0.0184 & 0.0327 & 0.0401 & 0.0338 & 0.0412 & 0.0400 & \textbf{0.0419} & \underline{0.0416} & 0.0382 & 0.0407 & 0.0397 & 0.0408$\text{}^{\ddagger}$ \\
& N@20 & 0.0243 & 0.0385 & 0.0471  & 0.0416 & 0.0483 & 0.0470 & 0.0496 & 0.0497 & 0.0482 & \underline{0.0504} & 0.0502 & \textbf{0.0509}$\text{}^{\dagger \ddagger}$ \\ \hline
\end{tabular}
}
\end{table*}
\textit{Item Denoising for Sequential Recommendation.} We also evaluate \method{} against three baselines that either explicitly remove noisy items or implicitly address item noise through denoising modeling. All these models are implemented based on the item-ID base recommendation model, such as Bert4Rec~\cite{Sun2019Bert4rec}. For methods that explicitly remove noisy items, DSAN~\cite{yuan2021dual} learns a target item embedding and applies an adaptively sparse transformation function to filter out noisy interactions. Similarly, HSD~\cite{zhang2022hierarchical} identifies and excludes noisy interactions by detecting inconsistencies with user interests, leveraging the sequential order of recorded interactions.
In contrast, DCRec~\cite{yang2023debiased} employs a debiased contrastive learning framework to implicitly mitigate the impact of noise.

\textbf{Implementation Details.} \method{} is implemented using OpenMatch\footnote{\url{https://github.com/OpenMatch/OpenMatch}}~\cite{liu2021openmatch,yu2023openmatch} and optimized using in-batch negatives. Both the item representation model in the semantic shortcut extractor and the recommendation module are initialized from the T5-base checkpoint\footnote{\url{https://huggingface.co/google-t5/t5-base}} in Huggingface Transformers~\cite{wolf2019huggingface}. The item representation model in the semantic shortcut extractor is trained using the Adam optimizer with a learning rate of 1e-4 and a batch size of 32. For the recommendation module, we follow the settings in \citet{liu2023text}, which is trained using the Adam optimizer with a learning rate of 1e-4, a warm-up proportion of 0.1, and a batch size of 8. We set the filtering threshold $\tau$ to 0.7, based on the hyperparameter sensitivity analysis reported in Appendix~\ref{app:param}. During inference, we excluded the last interaction from the user-interaction sequence when constructing the consistency graph, thereby avoiding potential data leakage. 

Additionally, we reproduce all item-ID based models and DWSRec~\cite{zhang2024dual} using implementations provided by RecBole\footnote{\url{https://recbole.io/}}~\cite{zhao2021recbole}, maintaining consistent settings across all models for fair comparison. For CoWPiRec and TASTE, we adhere to the default experimental settings as specified in their original papers. For HLLM, due to computational constraints, we ultilize TinyLlama-1.1B checkpoint\footnote{\url{https://huggingface.co/TinyLlama/TinyLlama-1.1B-intermediate-step-1431k-3T}}~\cite{zhang2024tinyllama} as the item-LLM component, and the user-LLM is implemented as a custom Llama model~\cite{touvron2023llama} with two hidden layers, trained from scratch using the following settings: learning rate of 1e-4, batch size of 2, training for 5 epochs, and a negative sampling ratio of 1024 negative samples per positive sample.

\section{Experimental Results}
In this section, we begin by evaluating the recommendation performance of \method{} and perform ablation studies to assess the effectiveness of the semantic shortcut extractor and various training strategies. Then, we analyze textual similarities among retained items to evaluate the effectiveness of the item representation learning method. Finally, several case studies are shown.

\subsection{Overall Performance}  
In this section, we first implement \method{} based on TASTE~\cite{liu2023text} to evaluate its recommendation performance. Subsequently, we implement \method{} with Bert4Rec~\cite{Sun2019Bert4rec} to evaluate its denoising capability by comparing it against various baselines.

\subsubsection{Recommendation Performance} As shown in Table~\ref{tab:overall}, we present the overall performance of \method{} and various recommendation baseline models. Overall, \method{} demonstrates its effectiveness by significantly outperforming all baselines across four datasets, leveraging TASTE~\cite{liu2023text} as its backbone architecture.

Among all baseline models, text-based recommendation models generally exhibit significantly stronger performance compared to item ID-based models. This superiority derives from the ability of PLMs and LLMs to effectively capture matching signals through text representations of users and items. Based on the TASTE backbone, \method{} integrates a semantic shortcut extractor to remove noisy items, utilizing these retained user-item interactions to train the recommendation module. The evaluation results reveal a significant improvement in the recommendation performance, highlighting its effectiveness in enabling the recommendation module to better model user behavior. Among various recommendation tasks, \method{} demonstrates much better performance on the Beauty and Sports datasets, while showing relatively weaker improvements on the Yelp dataset. This discrepancy may be attributed to the tendency of users to make more casual clicks driven by instant interest in these shopping scenarios.

\begin{table}[t]
\centering
\small
\caption{\label{tab:filteroverall}Denoising Performance of \method{}. To ensure a fair comparison, all denoising methods are implemented using Bert4Rec~\cite{Sun2019Bert4rec} as the backbone model.}
\resizebox{\linewidth}{!}{
\begin{tabular}{l|lrrrr}
\hline
\textbf{Dataset} & \textbf{Method} & \textbf{R@10} & \textbf{R@20} & \textbf{N@10} & \textbf{N@20} \\ \hline
\multirow{4}{*}{Beauty} & DSAN & 0.0215 & 0.0320 & 0.0103 & 0.0132 \\ 
~ & HSD & 0.0423 & 0.0645 & 0.0218 & 0.0262 \\ 
~ & DCRec & \underline{0.0609} & \underline{0.0869} & \underline{0.0317} & \underline{0.0387} \\ 
~ & \method{} & \textbf{0.0627} & \textbf{0.0904} & \textbf{0.0336} & \textbf{0.0406} \\ 
\hline
\multirow{4}{*}{Sports} & DSAN & 0.0107 & 0.0159 & 0.0051 & 0.0065 \\ 
~ & HSD & 0.0214 & 0.0317 & 0.0103 & 0.0133 \\ 
~ & DCRec & \underline{0.0262} & \underline{0.0407} & \underline{0.0139} & \underline{0.0175} \\ 
~ & \method{} & \textbf{0.0275} & \textbf{0.0428} & \textbf{0.0147} & \textbf{0.0185} \\ 
\hline
\multirow{4}{*}{Toys} & DSAN & 0.0172 & 0.0258 & 0.0086 & 0.0104 \\ 
~ & HSD & 0.0344 & 0.0517 & 0.0172 & 0.0209 \\ 
~ & DCRec & \underline{0.0443} & \underline{0.0660} & \underline{0.0243} & \underline{0.0296} \\ 
~ & \method{} & \textbf{0.0468} & \textbf{0.0691} & \textbf{0.0258} & \textbf{0.0314} \\ 
\hline
\multirow{4}{*}{Yelp} & DSAN & 0.0153 & 0.0218 & 0.0084 & 0.0116 \\ 
~ & HSD & 0.0326 & 0.0483 & 0.0165 & 0.0247 \\ 
~ & DCRec & \underline{0.0387} & \underline{0.0562} & \underline{0.0204} & \underline{0.0262} \\ 
~ & \method{} & \textbf{0.0398} & \textbf{0.0576} & \textbf{0.0211} & \textbf{0.0274} \\ 
\hline
\end{tabular}
}
\vspace{-1.5em}
\end{table}
\subsubsection{Denoising Effectiveness} We compare \method{} with three state-of-the-art denoising methods--DSAN~\cite{yuan2021dual}, HSD~\cite{zhang2022hierarchical}, and DCRec~\cite{yang2023debiased}--to evaluate its denoising effectiveness. The baselines are categorized into two groups--explicit denoising methods (DSAN and HSD) and implicit denoising methods (DCRec). All methods are implemented using the item ID-based recommendation model, Bert4Rec~\cite{Sun2019Bert4rec}. The evaluation results are shown in Table~\ref{tab:filteroverall}.

Among the baseline methods, DCRec achieves the best performance by modeling user behaviors based on raw user-item interactions and effectively mitigating noise through dynamically assigning weights to user-interacted items. However, this implicit denoising approach necessitates model-specific training, restricting our experiments to the Bert4Rec model used in constructing the DCRec framework. Benefiting from explicit denoising methods, \method{} can be applied to the item ID-based recommendation model and achieves a 4.7\% improvement over all denoising baseline models, highlighting its effectiveness in filtering noisy items. This consistent improvement with the item ID-based model underscores the potential of extending the filtering approach of \method{} to different architectures of recommendation models.

\begin{table}[t]
\centering
\small
\caption{\label{tab:ablation}Ablation Study of \method{}. ${\dagger}$ and ${\ddagger}$ denote statistically significant improvements over $\text{\method{} (w/o $M_\text{Filter}$)}^{\dagger}$ and $\text{\method{} (Zero-Shot)}^{\ddagger}$.}
\resizebox{\linewidth}{!}{
\begin{tabular}{l|lll}
\hline
\textbf{Dataset} & \textbf{Method} & \textbf{R@20} & \textbf{N@20} \\ \hline
\multirow{5}{*}{Beauty} & \method{} (Zero-Shot) & 0.1631 & 0.0697 \\
& w/o $M_\text{Filter}$ & 0.1550 & 0.0649 \\ \cdashline{2-4}
~ & \method{} & \textbf{0.1774}$\text{}^{\dagger \ddagger}$ & \textbf{0.0772}$\text{}^{\dagger \ddagger}$ \\
~ & w/o $M_\text{Rec}$ Training & 0.1655$\text{}^{\dagger}$ & 0.0727$\text{}^{\dagger}$ \\ 
~ & w/o $M_\text{Filter}$ Training  & \underline{0.1761}$\text{}^{\dagger \ddagger}$ & \underline{0.0762}$\text{}^{\dagger \ddagger}$ \\ \hline
\multirow{5}{*}{Sports} & \method{} (Zero-Shot) & 0.1014 & 0.0420 \\
~ & w/o $M_\text{Filter}$ & 0.0964 & 0.0421 \\ \cdashline{2-4}
~ & \method{} & \textbf{0.1082}$\text{}^{\dagger \ddagger}$ & \textbf{0.0496}$\text{}^{\dagger \ddagger}$ \\
~ & w/o $M_\text{Rec}$ Training & 0.1047$\text{}^{\dagger}$ & 0.0433$\text{}^{\dagger}$ \\ 
~ & w/o $M_\text{Filter}$ Training & \underline{0.1073}$\text{}^{\dagger \ddagger}$ & \underline{0.0457}$\text{}^{\dagger \ddagger}$ \\ \hline
\multirow{5}{*}{Toys} & \method{} (Zero-Shot) & 0.1797 & 0.0786 \\
~ & w/o $M_\text{Filter}$ & 0.1789 & 0.0780 \\ \cdashline{2-4}
~ & \method{} & \textbf{0.1896}$\text{}^{\dagger \ddagger}$ & \textbf{0.0826}$\text{}^{\dagger \ddagger}$ \\
~ & w/o $M_\text{Rec}$ Training & 0.1836$\text{}^{\dagger}$ & 0.0796$\text{}^{\dagger}$ \\ 
~ & w/o $M_\text{Filter}$ Training & \underline{0.1879}$\text{}^{\dagger \ddagger}$ & \underline{0.0802}$\text{}^{\dagger \ddagger}$ \\ \hline
\multirow{5}{*}{Yelp} & \method{} (Zero-Shot)  & 0.1158 & 0.0504 \\
~ & w/o $M_\text{Filter}$ & 0.1156 & 0.0502 \\ \cdashline{2-4}
~ & \method{} & \textbf{0.1173}$\text{}^{\dagger}$ & \textbf{0.0509}$\text{}^{\dagger}$ \\
~ & w/o $M_\text{Rec}$ Training & 0.1165 & 0.0504 \\ 
~ & w/o $M_\text{Filter}$ Training & \underline{0.1169}$\text{}^{\dagger}$ & \underline{0.0507}$\text{}^{\dagger}$ \\ \hline
\end{tabular}
}
\end{table}
\begin{table}[t]
\centering
\small
\caption{\label{tab:ablation2}Ablation Study on Different Filter Models in \method{}.}
\resizebox{\linewidth}{!}{
\begin{tabular}{l|lll}
\hline
\textbf{Dataset} & \textbf{Method} & \textbf{R@20} & \textbf{N@20} \\ \hline
\multirow{3}{*}{Beauty} & \method{} & \textbf{0.1774} & \textbf{0.0772} \\
& w/ Vanilla T5 $M_\text{Filter}$ & 0.1761 & 0.0762 \\
~ & w/ TASTE $M_\text{Filter}$ & \underline{0.1767} & \underline{0.0767} \\ \hline
\multirow{3}{*}{Sports} & \method{} & \textbf{0.1082} & \textbf{0.0496} \\
~ & w/ Vanilla T5 $M_\text{Filter}$ & 0.1073 & 0.0457 \\
~ & w/ TASTE $M_\text{Filter}$ & \underline{0.1077} & \underline{0.0471} \\ \hline
\multirow{3}{*}{Toys} & \method{} & \textbf{0.1896} & \textbf{0.0826} \\
~ & w/ Vanilla T5 $M_\text{Filter}$ & 0.1879 & 0.0802 \\
~ & w/ TASTE $M_\text{Filter}$ & \underline{0.1883} & \underline{0.0811} \\ \hline
\multirow{3}{*}{Yelp} & \method{} & \textbf{0.1173} & \textbf{0.0509} \\
~ & w/ Vanilla T5 $M_\text{Filter}$ & 0.1169 & 0.0507 \\
~ & w/ TASTE $M_\text{Filter}$ & \underline{0.1170} & \underline{0.0508} \\ \hline
\end{tabular}
}
\end{table}
\subsection{Ablation Study}
This section presents an ablation study to evaluate the contribution of the semantic shortcut extractor ($M_\text{Filter}$) and the effectiveness of our proposed training strategy in \method{}. 

Table~\ref{tab:ablation} shows the performance of several ablated versions of our model across four datasets. Comparing \method{} (w/o $M_\text{Filter}$) with \method{} (Zero-Shot), we observe a consistent improvement in recommendation performance across all datasets when a zero-shot $M_\text{Filter}$ is applied during inference. This highlights the inherent benefit of even an untrained semantic shortcut extractor in refining user-interacted items for recommendation. The impact of the semantic shortcut extractor becomes more evident when comparing \method{} (Zero-Shot) with \method{} (w/o $M_\text{Rec}$ Training). Training the $M_\text{Filter}$ module with our instruction-based approach results in significant performance gains, demonstrating the value of learning tailored item representations for extracting semantic shortcuts. In particular, \method{} (w/o $M_\text{Rec}$ Training) consistently outperforms \method{} (w/o $M_\text{Filter}$), highlighting that an effective semantic shortcut extractor is essential for improved recommendation accuracy.

Furthermore, the full \method{} model, which trains the recommendation module $M_\text{Rec}$ on the retained interactions, achieves the best performance across all datasets, with an average improvement of over 12\% compared to \method{} (Zero-Shot). This substantial gain underscores the importance of aligning the recommendation module with the retained data, enabling it to better capture user preferences and behaviors from the refined interaction history. The consistent statistically significant improvements of \method{} over both \method{} (w/o $M_\text{Filter}$) and \method{} (Zero-Shot) further validate the effectiveness of our complete model.

To evaluate the effectiveness of the $M_\text{Filter}$ module in \method{}, we compare it with two alternatives: a vanilla T5 $M_\text{Filter}$ initialized from T5-base without task-specific training, and a TASTE $M_\text{Filter}$ (trained on recommendation data).

As shown in Table~\ref{tab:ablation2}, \method{} achieves the best performance across all datasets, consistently surpassing both baselines. The vanilla T5 $M_\text{Filter}$ lags behind due to its lack of adaptation to recommendation data, while the TASTE-based variant, though improved, is still limited by its exposure to only a single dataset during training. In contrast, our approach benefits from balanced pre-training over multiple datasets, leading to more robust and transferable item representations. These results highlight the importance of both domain-specific and multi-dataset training for semantic shortcut extraction in recommendation.

\begin{figure}[t]
\centering 
\subfigure[Bert4Rec] { 
\label{fig:param:a} 
\includegraphics[width=0.45\linewidth]{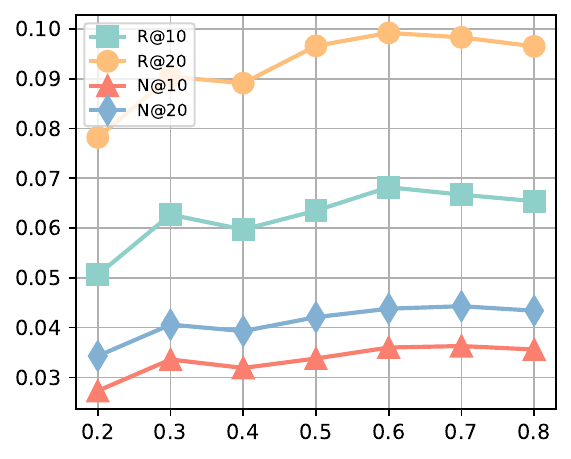}}
\subfigure[TASTE] { 
\label{fig:param:b}
\includegraphics[width=0.45\linewidth]{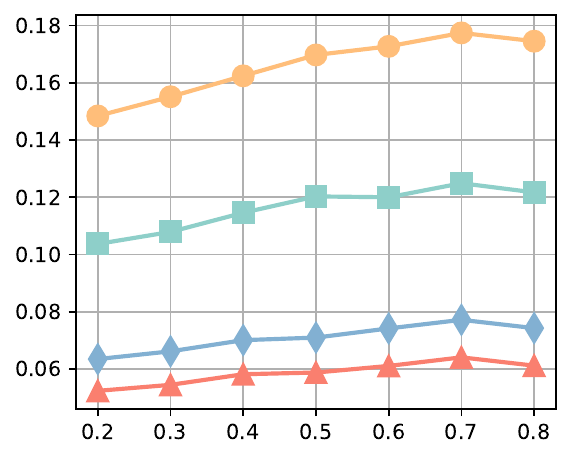}}
\caption{\label{fig:param}Hyperparameter Sensitivity Analysis. We conduct experiments on the threshold hyperparameter, $\tau$, with results derived from the Amazon Beauty dataset using the Bert4Rec~\ref{fig:param:a} and TASTE~\ref{fig:param:b} backbone models.}
\end{figure}
\subsection{Hyperparameter Sensitivity Analysis}\label{app:param}
We analyze the impact of the similarity threshold $\tau$, which is a key hyperparameter in the construction of the personal item graph described in Sec.\ref{model:denoising}, by varying $\tau$ from 0.2 to 0.8 while keeping all other experimental settings fixed. Figure~\ref{fig:param} shows the performance of Bert4Rec and TASTE across four evaluation metrics (Recall@10/20 and NDCG@10/20).

The results reveal a clear and consistent pattern for both backbone models. When $\tau$ is set too low (0.2 to 0.4), the semantic shortcut extractor becomes overly permissive, allowing many irrelevant or noisy items into the shortcut, which results in degraded performance. As $\tau$ increases to a moderate range (0.5 to 0.6), performance improves markedly, indicating that the shortcut extractor is now able to focus on truly relevant interactions and effectively filter out noise. For example, Bert4Rec achieves its best R@10 around $\tau=0.6$, and similar patterns are observed for all metrics and for the TASTE backbone.

Importantly, our \method{} framework consistently outperforms baseline models without semantic filtering across the entire range of $\tau$ values. This demonstrates the robustness and generalizability of our approach--even when the threshold is not optimally tuned, shortcut extraction leads to reliable improvements in recommendation quality.

\begin{figure}[t]
    \centering
    \subfigure[Raw User-Item Interaction Trained $M_\text{Rec}$.]{
        \label{fig:item:a}
        \includegraphics[width=0.45\linewidth]{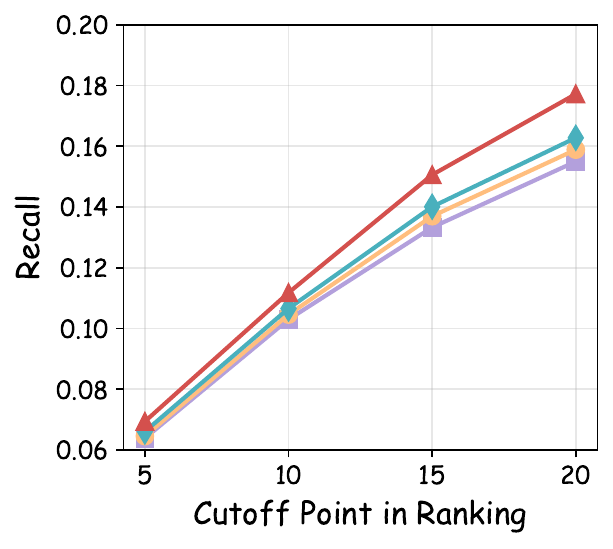}
    }
    \hfill
    \subfigure[$M_\text{Filter}$-Retained Items Trained $M_\text{Rec}$.]{
        \label{fig:item:b}
        \includegraphics[width=0.45\linewidth]{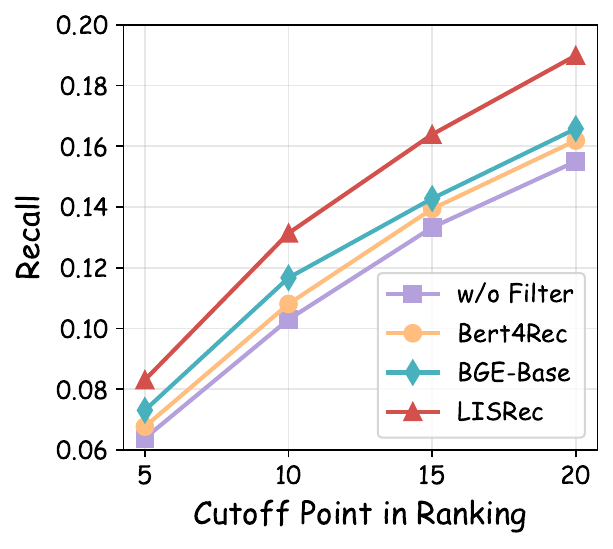}
    }
    \caption{\label{fig:item}Performance of Recommendation Models Using Different Semantic Shortcut Extraction Methods.}
\end{figure}
\subsection{Effectiveness of Different Item Representations for Shortcut Extraction}\label{result:training} 
To evaluate the effectiveness of the semantic shortcut extractor in \method{}, we perform a comparative analysis involving various item representation strategies. Specifically, we compare the full \method{} model with its variant without the semantic shortcut extractor (\method{} w/o Filter), as well as versions of \method{} equipped with different item representation methods within the semantic shortcut extractor. 

We implement the semantic shortcut extractor $M_\text{Filter}$ using three approaches: an ID-based recommendation model, Bert4Rec~\cite{Sun2019Bert4rec}, and a dense retriever, BGE-Base~\cite{bge_embedding}, which serve as baselines for evaluating representation effectiveness.

\begin{figure}[t]
    \centering
    \subfigure[Average Similarity Scores among $M_\text{Filter}$-Retained Items.]{
        \label{fig:sim:a}
        \includegraphics[width=0.45\linewidth]{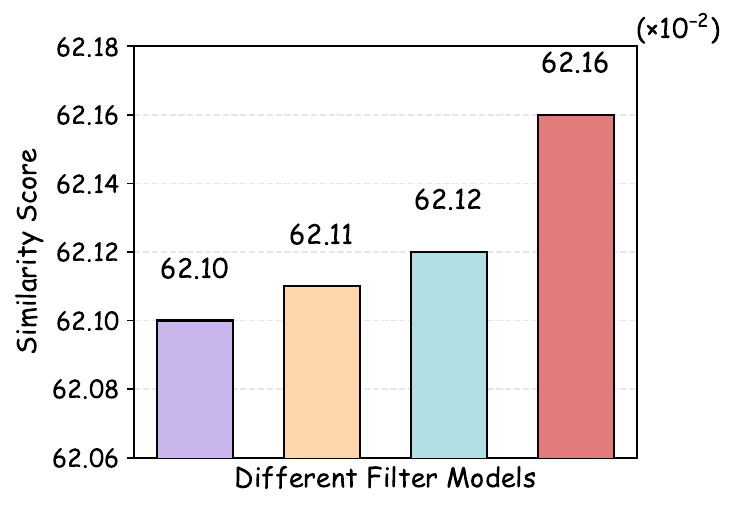}
    }
    \hfill
    \subfigure[Similarity Scores between $M_\text{Filter}$-Retained Items and the Target Items.]{
        \label{fig:sim:b}
        \includegraphics[width=0.45\linewidth]{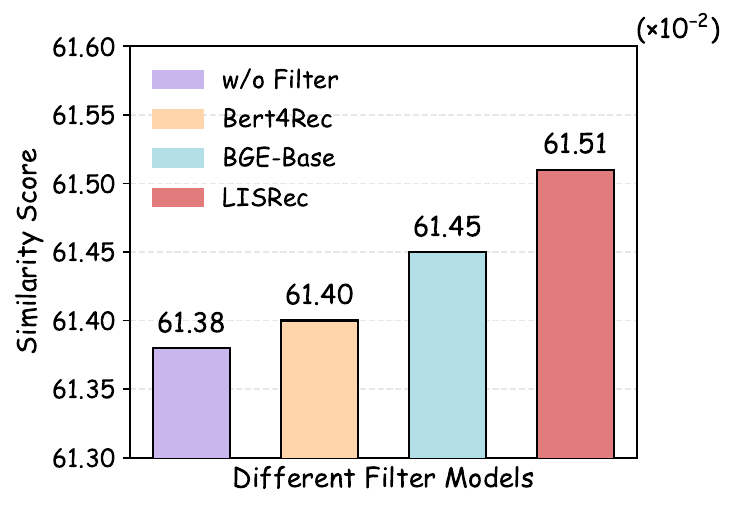}
    }
    \caption{Effectiveness of \method{} in Learning Item Representations for Semantic Shortcut Extraction. Figure~\ref{fig:sim:a} and Figure~\ref{fig:sim:b} illustrate the item similarities among $M_\text{Filter}$-retained items and the ground truth items.}
    \label{fig:sim}
\end{figure}
As illustrated in Figure~\ref{fig:sim}, we analyze the text similarity among the items retained by $M_\text{Filter}$ using the BGE-Base model to compute pairwise similarity scores. In Figure~\ref{fig:sim:a}, we report the average similarity among the items selected by different semantic shortcut extraction strategies. The results show that all methods increase the internal similarity among selected items. Notably, when using \method{}, the extracted item subset achieves the highest average similarity, indicating stronger semantic alignment within the shortcut.

Furthermore, Figure~\ref{fig:sim:b} presents the average similarity between the $M_\text{Filter}$-retained user-interacted items and the ground-truth target item. Among all methods, \method{} yields the highest similarity, demonstrating its effectiveness in maintaining items that are semantically aligned with the target item that users need.

As shown in Figure~\ref{fig:item}, we further evaluate the impact of different shortcut extraction methods on overall recommendation performance. Figures~\ref{fig:item:a} and~\ref{fig:item:b} report the performance of the recommendation module $M_\text{Rec}$ when trained on either the raw user-item interaction set $\mathcal{H}$ or the shortcut subset $\mathcal{H}_\text{Shortcut}$ extracted by $M_\text{Filter}$. In both cases, \method{} consistently outperforms other item representation models, with particularly strong gains at larger ranking cutoffs. This demonstrates its effectiveness in enabling the recommendation model to focus on the most relevant items, thereby better capturing stable user preferences. The advantage of \method{} is attributed to its PLM-based ability to capture item text semantics and to identify semantically meaningful patterns in user-item interactions. After training the recommendation module $M_\text{Rec}$ using the shortcut-extracted interaction histories, \method{} achieves substantial performance improvements, further highlighting the effectiveness of semantic shortcut extraction.

\begin{table*}[t]
\centering
\small
\renewcommand\arraystretch{1.5}
\caption{\label{tab:case1}Case Studies. We present cases from two datasets, Amazon Beauty and Yelp. Different colors are highlighted to distinguish preference types: \sethlcolor{lightgreen}\hl{Blue} represents majority preference, and \sethlcolor{lightred}\hl{Orange} represents discrete interaction.}
\resizebox{\textwidth}{!}{
\begin{tabular}{l}
\hline
\multicolumn{1}{l}{\textit{\textbf{Case \#1 in Amazon Beauty}}} \\ \hline
\textbf{Node\_ID:} $n_1$\qquad\qquad\textbf{Name:} Bare Escentuals bareMinerals Purely Nourishing Moisturizer: Combination Skin \\
\textbf{Description:} ... \sethlcolor{lightgreen}\hl{Hydrate skin} with this lightweight, remarkably effective \sethlcolor{lightgreen}\hl{moisturizer} for improved firmness, elasticity ... \\ \cdashline{1-1}
\textbf{Node\_ID:} $n_2$\qquad\qquad\textbf{Name:} Ole Henriksen Truth Creme Advanced Hydration, \sethlcolor{lightgreen}\hl{1.7} Fluid Ounce \\
\textbf{Description:} ... An effective and groundbreaking anti-aging \sethlcolor{lightgreen}\hl{moisturizer} contains \sethlcolor{lightgreen}\hl{advanced hydration complexity} that deliver ... \\ \cdashline{1-1}
\textbf{Node\_ID:} $n_3$\qquad\qquad\textbf{Name:} VALENTINO VALENTINA by Valentino for WOMEN: EAU DE PARFUM SPRAY \sethlcolor{lightgreen}\hl{1.7} OZ \\
\textbf{Description:} ... Valentina By Valentino \sethlcolor{lightgreen}\hl{1.7} Oz Eau De \sethlcolor{lightred}\hl{Parfum Spray} ... \\ \cdashline{1-1}
\textbf{Node\_ID:} $n_4$\qquad\qquad\textbf{Name:} Boscia Oil-Free Nightly Hydration, 1.4-Fluid Ounce \\
\textbf{Description:} ... An oil-free nightly \sethlcolor{lightgreen}\hl{moisturizer} that provides \sethlcolor{lightgreen}\hl{weightless hydration}, combats excess oil and refines the look of ... \\ \hline
Similarity: \sethlcolor{lightgreen}\hl{$S_{1,2}=0.4441$}, \sethlcolor{lightred}\hl{$S_{1,3}=0.2527$}, \sethlcolor{lightgreen}\hl{$S_{1,4}=0.4608$}, \sethlcolor{lightred}\hl{$S_{2,3}=0.2125$}, \sethlcolor{lightgreen}\hl{$S_{2,4}=0.4477$}, \sethlcolor{lightred}\hl{$S_{3,4}=0.2049$} \\
Add Edge: $(n_1,n_2)$, $(n_1,n_4)$, $(n_2,n_4)$ \\
The maximum connected subgraph $C_\text{max}$ is \sethlcolor{lightgreen}\hl{$\left\{1,2,4\right\}$}, so that the filtered interaction sequence $\mathcal{H}_\text{Shortcut}$ is: \sethlcolor{lightgreen}\hl{$\left\{1,2,4\right\}$} \\ \hline
\multicolumn{1}{l}{\textit{\textbf{Case \#2 in Yelp}}} \\ \hline
\textbf{Node\_ID:} $n_1$\qquad\qquad\textbf{Address:} \sethlcolor{lightgreen}\hl{Las Vegas, NV}\qquad\qquad\textbf{Name:} Maggiano's Little Italy \\
\textbf{Description:} ... \sethlcolor{lightgreen}\hl{Nightlife}, \sethlcolor{lightgreen}\hl{Restaurants}, Italian, Bars ... \\ \cdashline{1-1}
\textbf{Node\_ID:} $n_2$\qquad\qquad\textbf{Address:} \sethlcolor{lightgreen}\hl{Las Vegas, NV}\qquad\qquad\textbf{Name:} Lazy Dog \sethlcolor{lightgreen}\hl{Restaurant} \& Bar \\
\textbf{Description:} ... \sethlcolor{lightgreen}\hl{Nightlife}, \sethlcolor{lightgreen}\hl{Food}, Beer, Wine \& Spirits, Pizza, Burgers ... \\ \cdashline{1-1}
\textbf{Node\_ID:} $n_3$\qquad\qquad\textbf{Address:} \sethlcolor{lightgreen}\hl{Las Vegas, NV}\qquad\qquad\textbf{Name:} Spay \& Neuter Center of Southern Nevada \\
\textbf{Description:} ... \sethlcolor{lightred}\hl{Veterinarians}, Pets, \sethlcolor{lightred}\hl{Pet Services} ... \\ \cdashline{1-1}
\textbf{Node\_ID:} $n_4$\qquad\qquad\textbf{Address:} \sethlcolor{lightgreen}\hl{Las Vegas, NV}\qquad\qquad\textbf{Name:} Weera Thai \sethlcolor{lightgreen}\hl{Restaurant} - Sahara \\
\textbf{Description:} ... \sethlcolor{lightgreen}\hl{Restaurants}, Beer, Wine \& Spirits, \sethlcolor{lightgreen}\hl{Food}, Vegetarian, ..., Seafood, Bars, \sethlcolor{lightgreen}\hl{Nightlife} ... \\ \hline
Similarity: \sethlcolor{lightgreen}\hl{$S_{1,2}=0.5926$}, \sethlcolor{lightred}\hl{$S_{1,3}=0.4699$}, \sethlcolor{lightgreen}\hl{$S_{1,4}=0.5121$}, \sethlcolor{lightred}\hl{$S_{2,3}=0.4865$}, \sethlcolor{lightgreen}\hl{$S_{2,4}=0.5150$}, \sethlcolor{lightred}\hl{$S_{3,4}=0.3932$} \\
Add Edge: $(n_1,n_2)$, $(n_1,n_4)$, $(n_2,n_4)$ \\
The maximum connected subgraph $C_\text{max}$ is \sethlcolor{lightgreen}\hl{$\left\{1,2,4\right\}$}, so that the filtered interaction sequence $\mathcal{H}_\text{Shortcut}$ is: \sethlcolor{lightgreen}\hl{$\left\{1,2,4\right\}$} \\ \hline
\end{tabular}}
\end{table*}
\begin{figure}[t]
    \centering
    \subfigure[Similarity Scores between User Representation and Target Items.]{
        \label{fig:emb:a}
        \includegraphics[width=0.45\linewidth]{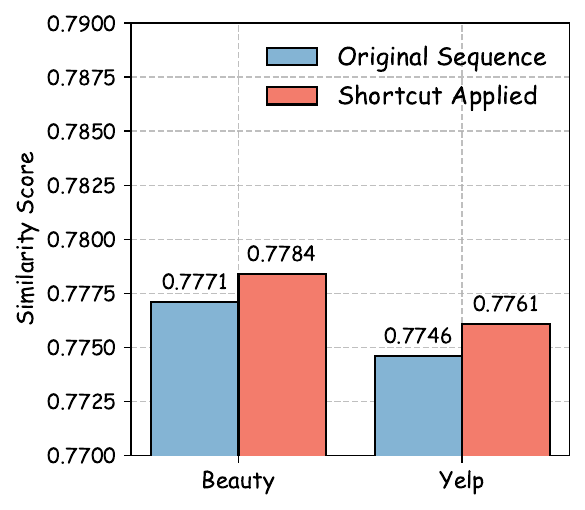}
    }
    \hfill
    \subfigure[Embedding Visualization of Items using \method{}.]{
        \label{fig:emb:b}
        \includegraphics[width=0.45\linewidth]{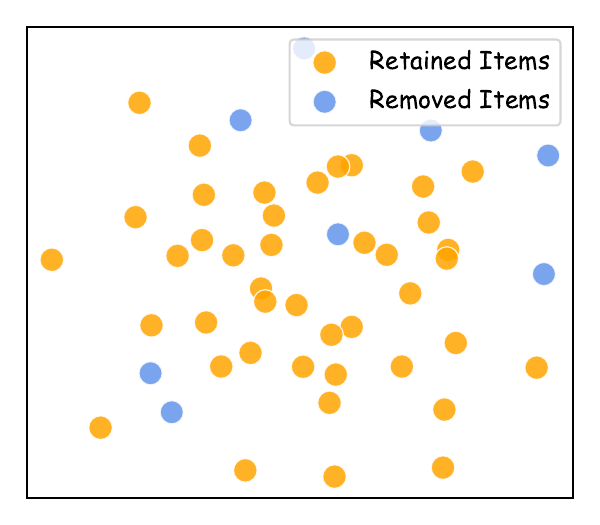}
    }
    \caption{
        Effectiveness of Semantic Shortcut Extraction in Enhancing the Recommendation Module. Figure~\ref{fig:emb:a} and Figure~\ref{fig:emb:b} show the improvement of user and item representations after extracting shortcuts.
    }
    \label{fig:emb}

\vspace{-1em}
\end{figure}
\subsection{Effectiveness of Shortcut Extraction in Enhancing the Recommendation Module}
As illustrated in Figure~\ref{fig:emb}, we investigate how the shortcut-retained items contribute to improving the recommendation module during training.

For each user, we first compute the embedding similarity between the user representation and the corresponding ground-truth item, under both the original and shortcut-applied training settings. Figure~\ref{fig:emb:a} shows that applying shortcut extraction during training consistently increases this similarity. This finding confirms that shortcut extraction strengthens the semantic alignment between user and target item representations, leading to more faithful modeling of user intent.

Furthermore, we randomly select a user and then use t-SNE to visualize item embeddings. As shown in Figure~\ref{fig:emb:b}, \method{} is able to better distinguish the removed items by mapping them to the periphery of the embedding space, effectively treating them as outliers. This observation highlights the effectiveness of \method{} in leveraging denoising signals as prior knowledge to guide the recommendation module. By doing so, it facilitates a more discriminative representation space, where retained and removed items are clearly separated through contrastive training.

\subsection{Case Studies}\label{app:case}
In Table~\ref{tab:case1}, we present two examples from the Amazon Beauty and Yelp datasets to evaluate the effectiveness of \method{}. For each dataset, we randomly select a user and showcase how semantic shortcut extraction operates on their interaction history. In the case studies, we detail the procedure of constructing the personalized item interaction graph and extracting the maximal preference-consistent subgraph, which serves to filter out noisy interactions.

As shown in the first case from the Amazon Beauty dataset, the user primarily interacts with face cream products focused on moisturizing, such as \textit{``Bare Escentuals bareMinerals Purely Nourishing Moisturizer''} and \textit{``Ole Henriksen Truth Creme Advanced Hydration''}. However, the user accidentally clicks on a product--\textit{``VALENTINO VALENTINA by Valentino for WOMEN: EAU DE PARFUM SPRAY 1.7 OZ''}--due to their instant interests, which are significantly different in terms of both product types and attributes. In detail, while other items in the sequence emphasize \textit{``hydrating''} and \textit{``moisturizing''}, the \textit{``Parfum Spray''} is unrelated to these preferences. When \method{} constructs the user-interacted item graph, the \textit{``Parfum Spray''} product fails to form edges with other items, indicating that this interaction is likely an outlier and should be excluded from user-interacted items. Consequently, the denoised item sequence $\mathcal{H}_\text{Shortcut}$ includes only products aligned with the consistent preferences of a specific user, such as moisturizing creams, enabling the recommendation module to better model user behavior from the user-item interactions.

In the second case, we present a specific user from the Yelp dataset who visits various locations. This user primarily interacts with places characterized by attributes such as \textit{``Nightlife''}, \textit{``Restaurants''}, and \textit{``Bars''}, demonstrating a consistent pattern of preferences. However, one of the interacted locations--the \textit{``Spay \& Neuter Center of Southern Nevada''}--is a pet hospital, which is distinct from the user-consistent preference for nightlife and dining. Although this pet hospital is situated on the same street as other frequently visited locations, it likely represents an occasional visit rather than a reflection of the general behavior of the user. \method{} assigns a lower similarity score between this noisy item and other user-interacted items, effectively excluding the noisy item from the sequence of user-interacted history. This enables the recommendation model to focus on more representative items, such as \textit{``Maggiano's Little Italy''} and \textit{``Weera Thai Restaurant''}.

\section{Conclusion}
This paper introduces the Learned Item Shortcuts for Recommendation System (\method{}), a new framework that models stable user preferences by identifying dominant item subsets within a latent semantic space. \method{} leverages universal item representations to construct semantically meaningful shortcuts in user interaction histories. 
Our experiments on Yelp and Amazon Product datasets show that \method{} consistently outperforms baseline and previous text-based models. The results demonstrate that instruction-tuned universal item representations enable more robust shortcut selection and lead to more accurate, preference-consistent recommendations.



\begin{acks}
This work is partly supported by the National Natural Science Foundation of China (No. 62206042 and No. 62461146205). This work is also supported by the AI9Stars community.
\end{acks}

\clearpage
\bibliographystyle{ACM-Reference-Format}
\bibliography{custom}

\newpage
\appendix
\balance

\section{Appendix}
\label{sec:appendix}
\subsection{License}\label{app:license}
We show the licenses of the datasets that we use. Yelp\footnote{\url{https://www.yelp.com/dataset}} uses Apache License 2.0, while Amazon Product shows its terms of use on website\footnote{\url{http://jmcauley.ucsd.edu/data/amazon/}}. All of these licenses and agreements allow their data for academic use.

\subsection{Additional Experimental Details of Data Preprocessing}\label{app:data}
In our experiments, we employ Recbole~\cite{zhao2021recbole} for preprocessing the data for sequential recommendation modeling, maintaining the same experimental settings as prior works~\cite{Zhou2020s3, Xie2022DIF-SR, liu2023text, chen2024hllm}. Specifically, each user-item interaction sequence $\{v_1, \dots, v_t\}$ is processed using the leave-one-out strategy~\cite{Xie2022DIF-SR, chen2018sequential, Sun2019Bert4rec, Zhou2020s3} to construct the training, development, and testing sets. In detail, the testing set predicts the $t$-th item $v_t$ based on the previously interacted items $v_{1:t-1}$ of the user, while the development set predicts the $t-1$-th item $v_{t-1}$ given the interactions $v_{1:t-2}$. The training set predicts $v_i$ using $v_{1:i-1}$, where $1 < i < t-1$. Items and users with fewer than five interactions are filtered out, and all user-item interactions are treated as implicit feedback~\cite{chen2018sequential, Sun2019Bert4rec, Zhou2020s3, Xie2022DIF-SR, liu2023text, chen2024hllm}. The data statistics are shown in Table~\ref{tab:dataset}.

\begin{table}[t]
\centering
\small
\caption{\label{tab:dataset}Statistics of the Datasets Used for Modeling the Sequential Recommendation Module ($M_\text{Rec}$).}
\resizebox{\linewidth}{!}{
\begin{tabular}{l|rrr|rrr}
\hline 
\multirow{2}{*}{\textbf{Dataset}} &  \multicolumn{3}{c|}{\textbf{Data Information}} &  \multicolumn{3}{c}{\textbf{Split}}\\
&\textbf{\#Users} & \textbf{\#Items} & \textbf{\#Actions} & \textbf{Train} & \textbf{Dev} & \textbf{Test}  \\ \hline
Beauty & 22,363 & 12,101 & 198,502 & 131,413 & 22,363 & 22,363 \\
Sports & 35,598 & 18,357 & 296,337 & 189,543 & 35,598 & 35,598 \\
Toys & 19,412 & 11,924 & 167,597 & 109,361 & 19,412 & 19,412 \\
Yelp & 30,499 & 20,068 & 317,182 & 225,685 & 30,499 & 30,499 \\
\hline
\end{tabular}
}
\end{table}
\begin{table}[t]
\centering
\small
\caption{\label{tab:dataset1}Statistics of Training Data Used in the Filter Module ($M_\text{Filter}$). The data is used for optimizing item representations to estimate their similarities.}
\begin{tabular}{l|rrr|rr}
\hline
\multirow{2}{*}{\textbf{Dataset}} & \multicolumn{3}{c|}{\textbf{Data Information}} & \multicolumn{2}{c}{\textbf{Split}} \\
~ & \textbf{\#Users} & \textbf{\#Items} & \textbf{\#Actions} & \textbf{Train} & \textbf{Dev} \\ \hline 
Beauty & 22,363 & 12,101 & 198,502 & 225,685 & 19,412 \\ 
Sports & 35,598 & 18,357 & 296,337 & 225,685 & 19,412 \\ 
Toys & 19,142 & 11,924 & 167,597 & 225,685 & 19,412 \\ 
Yelp & 30,499 & 20,068 & 317,182 & 225,685 & 19,412 \\ \hline 
Total & 107,602 & 62,450 & 979,618 & 902,740 & 77,648 \\ \hline
\end{tabular}
\end{table}
Then, we construct the dataset for training the language model to represent items for estimating their similarities (Eq.~\ref{eq:sij}). We start with the training and development sets used in sequential recommendation modeling and further process them for the Next Item Prediction (NIP) task. We first resample user-item interaction sequences from different datasets, namely Beauty, Sports, Toys, and Yelp. Then, we balance the number of data points to construct the training and development sets. This approach mitigates the risk of overfitting caused by data imbalance in different recommendation tasks. 

As shown in Table~\ref{tab:dataset1}, the data statistics for both the training and development sets used in item representation learning are presented.

Following previous studies~\cite{Xie2022DIF-SR, liu2023text}, we conduct evaluations in the full ranking testing scenario~\cite{dallmann2021case,krichene2022sampled}, where items are ranked across the entire item set rather than sampling subsets. This evaluation set mitigates the inconsistencies often observed in sampled evaluations, offering a more realistic and comprehensive assessment of recommendation performance.

\subsection{Theoretical Analysis of Stable User Preference Modeling via \method{}}\label{app:proof}
\textbf{Preliminary.} Let $\mathcal{H}=\{v_1,\dots,v_n\}$ be the full interaction history of a user. The consistent user preference set $P$ can be represented:
\begin{equation}
\small
P\subseteq\mathcal{H}, \lvert P\rvert=m,
\end{equation}
and the remaining items set can be denoted:
\begin{equation}
\small
N=\mathcal{H}\setminus P, \lvert N\rvert=n-m.
\end{equation}
Then, we denote the user interaction graph as an undirected graph:
\begin{equation}
\small
G_{\mathcal{U}}=(\mathcal{H},\mathcal{E}).
\end{equation}
Here
$(v_i, v_j) \in \mathcal{E}\Longleftrightarrow\cos(\vec v_i, \vec v_j\bigr)=S_{ij} \ge \tau$,
where the fixed threshold $\tau \in (0,1)$. In the user interaction graph $G_{\mathcal{U}}$, if every pair of nodes in a set can be connected by a path, then this set of nodes is called a ``connected component''. We denote the collection of all connected components as:
\begin{equation}
\small
\mathcal{C} = \{C_1, C_2, \dots, C_\ell\},
\end{equation}
where each connected component $C_k\subseteq\mathcal{H}$ and $C_i\cap C_j=\emptyset$ for all $ i\ne j$. Then, we denote
\begin{equation}
\small
C_\text{max}=\mathrm{arg}\max_{1\leq k\leq\ell}\lvert C_k\rvert
\end{equation}
as the connected component with the maximum number of nodes. Above all, our main claim is that the maximum connected component $C_\text{max}$ is exactly the user preference set $P$.

\textbf{Assumptions.} We need the following three weakest conditions to support the subsequent derivation.

(i) The preference nodes are mutually reachable in $G_\mathcal{U}$:
\begin{equation}
\small
\forall v_i, v_j \in P: v_i \sim_\tau v_j
\end{equation}

(ii) There is no edge between any noise node in $N$ and any preferred node in $P$: 
\begin{equation}
\small
\forall v_i\in P, v_j\in N: S_{ij}<\tau;
\end{equation}

(iii) Even if noise nodes have high similarity and form small clusters, their scale will not exceed that of the preference set $P$:
\begin{equation}
\small
\forall \{C_1^N,\ldots,C_q^N\}\subseteq\mathcal{C}: \max_{1\leq r\leq q}\lvert C_r^N\rvert\mathrm{<}m.
\end{equation}

\textbf{Lemmas.} The following three lemmas correspond to assumptions (i)-(iii), respectively.

(1) \textit{The preference set induces a connected component.} By assumption (i), every pair of nodes in $P$ can reach each other through edges whose similarities exceed $\tau$, hence all nodes of $P$ lie in the same connected component of $G_\mathcal{U}$. Moreover, by assumption (ii), no node in $N$ is adjacent to any node in $P$, so that component contains only the nodes of $P$.

(2) \textit{Noise nodes and preference nodes are not in the same component.} By assumption (ii), $\forall v_i \in P, v_j \in N$, then $S_{ij} < \tau$, which implies that there is no direct edge between $v_i$ and $v_j$. If a noise node and a preferred node are in the same connected component, there must exist a path between the two nodes. However, as long as there is one transition from a noise node to a preferred node, it will violate assumption (ii). Therefore, such a path cannot exist. Consequently, all noise nodes and the connected component belong to different components.

(3) \textit{There is an upper bound on the scale of the noise cluster.} By assumption (iii), any connected component $C_r^N$ composed solely of noise nodes satisfies $\rvert C_r^N \rvert \mathrm{<}m$.

\textbf{Proof of Proposition.} We prove the two directions $v\in P\iff v\in C_{\max}$.

\textit{If $v\in P$, then $v\in C_{\max}$.} By Lemma (1) and Lemma (3), any two nodes in $P$ are connected by an edge, so that $P$ itself forms a connected component, which we denote by $C_P$, and $\lvert C_P\rvert = \lvert P\rvert = m$. Then, each noise node has no edges with $P$, and there are no edges between noise nodes. So all noise nodes are either isolated or form small components with size less than $m$. Therefore, among all components, $\lvert C_P\rvert$ is the largest. Hence $C_{\max}=C_P$, and thus $P \subseteq C_\text{max}$.

\textit{If $v\in C_{\max}$, then $v\in P$.} Suppose, for contradiction, there exists $v_k\in N\cap C_{\max}$. Since $C_{\max}=C_P$, $v_k$ is connected to some $v_i\in P$, and there must be a path passing through several edges. However, by Lemma (2) and Lemma (3), there is no noise node connected to any node in $P$, and there are no edges between noise nodes. So, it is impossible to construct a path from $v_k$ to $v_i$. Therefore, noise nodes cannot appear in $C_{\max}$. Hence, $C_{\max} \subseteq P$.

Thus, we can prove that $C_\text{max}=P$.

\end{document}